\documentclass[10pt,letterpaper,twocolumn,aps,showpacs,preprintnumbers,amsmath,amssymb,nofootinbib,prl]{revtex4-2}

\usepackage{graphicx}% Include figure files
\usepackage{dcolumn}% Align table columns on decimal point
\usepackage{bm}% bold math
\usepackage[utf8]{inputenc}
\usepackage[T1]{fontenc}

\usepackage{color}
\usepackage{amsmath}
\usepackage{amssymb}
\usepackage[unicode=true,pdfusetitle, bookmarks=true,bookmarksnumbered=false,bookmarksopen=false,
 breaklinks=false,pdfborder={0 0 0},pdfborderstyle={},backref=false,colorlinks=true]
 {hyperref}
\hypersetup{
 pdfborderstyle={},colorlinks,linkcolor=red,citecolor=blue}

\makeatletter
\usepackage{setspace}
\usepackage{mathrsfs}
\usepackage{siunitx}
\usepackage{soul}

\makeatother
\begin{document}
\title[\texorpdfstring{$\quad$}{}]{Light-induced fictitious magnetic fields for quantum storage in cold atomic ensembles}

\author{Jianmin Wang\texorpdfstring{$^{1,2,\ast}$}{}, Liang Dong\texorpdfstring{$^{1,3,\ast}$}{}, Xingchang Wang\texorpdfstring{$^{1,2}$}{}, Zihan Zhou\texorpdfstring{$^{1,2}$}{}, Ying Zuo\texorpdfstring{$^{1,2,\#}$}{}, Georgios A. Siviloglou\texorpdfstring{$^{1,2,\dagger}$}{}, and J. F. Chen\texorpdfstring{$^{1,2,3,\ddagger}$}{}} 

\affiliation{\mbox{\texorpdfstring{$^{1}$}{}Shenzhen Institute for Quantum Science and Engineering}\\
\mbox{Southern University of Science and Technology, Shenzhen, 518055, China}\\
\mbox{\texorpdfstring{$^{2}$}{}International Quantum Academy, Shenzhen, 518048, China}\\ 
\mbox{\texorpdfstring{$^{3}$}{}Department of Physics, Southern University of Science and Technology, Shenzhen, 518055, China}\\
\texorpdfstring{$^{\ast}$}{}These authors contributed equally to the work.\\
\texorpdfstring{$^{\#}$}{}zuoying@iqasz.cn
\texorpdfstring{$^{\dagger}$}{}siviloglouga@sustech.edu.cn
\texorpdfstring{$^{\ddagger}$}{}chenjf@sustech.edu.cn}

\date{\today}

\begin{abstract}
In this work, we have demonstrated that optically generated fictitious magnetic fields can be utilized to extend the lifetime of quantum memories in cold atomic ensembles. All the degrees of freedom of an AC Stark shift such as polarization, spatial profile, and temporal waveform can be readily controlled in a precise manner. Temporal fluctuations over several experimental cycles, and spatial inhomogeneities along a cold atomic gas have been compensated by an optical beam. The advantage of the use of fictitious magnetic fields for quantum storage stems from the speed and spatial precision that these fields can be synthesized. Our simple and versatile technique can find widespread application in coherent pulse and single-photon storage in any atomic species.

\end{abstract}

\maketitle

%\tableofcontents

%\section{Introduction} %Do not worry about the sections now they are just for convenience
\textit{Introduction.} Storing and retrieving information in network nodes are crucial building elements for diverse applications ranging from long-distance quantum communication~\cite{Duan2001, Azuma2023Quantum} to large-scale quantum computation and simulation~\cite{RevModPhys.79.135}. Quantum storage of single photons as well as memories for coherent light have been realized in a plethora of physical systems ranging from single atoms in optical cavities~\cite{Specht2011Singleatom}, to hot atomic vapors~\cite{Phillips2001Storage} and Mott insulators of quantum degenerate gases~\cite{Schnorrberger2009Electromagnetically}, and from ion doped crystals~\cite{Longdell2005Stopped} to optical microresonators~\cite{Fiore2011Storing}. Cold atomic ensembles, in particular, can be ideal memory platform not only for leading to the highest efficiency storage in record~\cite{Hsiao2018Highly, Vernaz-Gris2018Highlyefficient} but also enabling storage of diverse types of quantum states~\cite{Wang2019Efficient,parniak2017wavevector,PhysRevLett.114.050502,PhysRevLett.124.210504}.

In recent years, while considerable progress has been made for the storage efficiency of absorptive quantum memories~\cite{Liu2021, Azuma2023Quantum, Sangouard2011} based on elongated cold atomic ensembles~\cite{Vernaz-Gris2018Highlyefficient,Hsiao2018Highly,Wang2019Efficient,Cao2020Efficient} a critical challenge still persists: the need for longer storage lifetimes. Up to now, quantum qubits or photonic entangled states with that exhibit storage efficiencies higher than 80\%, suffer from short storage lifetimes, typically below $\SI{100} {\micro\second}$ that is the threshold for quantum communication within metropolitan areas~\cite{Azuma2023Quantum, Wang2019Efficient,Cao2020Efficient}. 

Several mechanisms lead to the degradation of stored quantum states in an atomic ensemble over time. The influence of the Doppler broadening and motional dephasing of the stored spin waves can be reduced by increasing the wavelength of the spin waves via perfect phase matching between optical fields~\cite{Zhao2009Millisecond}, cooling the atomic gases to sub-Doppler temperatures~\cite{Saglamyurek2021Storing}, and further limiting the atomic motion by the means of optical lattices~\cite{Zhao2009Longlived}. However, even with minimized atomic motion atomic ensembles still suffer from decoherence caused by ambient magnetic fields, and therefore magnetically insensitive clock transitions, e.g., ($m_F,m_{F'}$)=(0,0) can be chosen~\cite{Ye2022Longlived}. Together with dynamical decoupling, which is efficient for dephasing due to inhomogeneous broadening, the storage lifetime can be extended to seconds, but the above operations generally result in lower atom numbers, and thus retrieval efficiencies in the order of $10\%$~\cite{Dudin2013Light,Rui2015Operating}.

To achieve both appreciable storage efficiency and lifetime, precise control of the residual magnetic fields on the atomic ensemble is essential. The deleterious effects of the external magnetic fields are typically addressed by active stabilization from compensation current-carrying coils~\cite{10.1063/1.5080093,Xu2019Ultralow,Hesse2021Direct}. However, to reach high optical densities and therefore sufficient storage efficiency, the atomic clouds must be more than 2 cm in length, and the spatial gradient of magnetic field caused by the compensation coils is then detrimental. Meter-scale compensations coils generate relatively smooth fields and therefore cannot compensate for inhomogeneities in the scale of millimeters. In addition, they are typically rather slow with activation times in the order of several milliseconds. Therefore, compensating for these spatial inhomogeneities in such timescales could benefit from a complementary experimental approach. One promising path to address these challenges involves exploiting the AC Stark effect~\cite{cohen1972experimental}. The AC Stark effect was originally introduced to describe the energy shifts on an atomic level induced by time-varying electric fields, and it is caused by the tendency of the atomic dipoles to align with these light-induced oscillating electric fields~\cite{AutlerTownes1955, CohenTannoudji1998}. As was demonstrated recently, AC Stark shifts of spatially imperfect optical beams can have a direct influence on free-induction decay signal (FID)~\cite{Leszczynski2018Spatially}. AC Stark shifts have emerged as a powerful tool in fundamental science and in quantum technology research with widespread application in atomic spectroscopy, laser cooling and trapping, spin wave manipulation~\cite{Chaudhury2009Quantum, Parniak2019Quantum, Leszczynski2018Spatially}, and even in single-spin addressing~\cite{Weitenberg2011} among many others.

In this Letter, we demonstrate the efficacy of precisely controlled fictitious magnetic fields, generated by AC Stark shifts~\cite{cohen1972experimental}, for extending the lifetime of quantum memories in cold atomic ensembles. 
Our approach harnesses the full flexibility of AC Stark shift beams, allowing precise engineering of polarization, spatial profile, and temporal waveform to compensate temporal fluctuations and spatial inhomogeneities with magnetic origin in an atomic ensemble. We demonstrate experimentally that the vector components of the AC Stark shifts can act on quantum storage equivalently to standard magnetic fields. Notably, with high speed and spatial precision, the fictitious magnetic field generated via non-destructive configurable AC Stark potentials offers a versatile method to manipulate atomic coherence in diverse physical systems such as atoms and ion-doped solids in various settings such as processors based on spin waves, gradient echo configurations, and even atomic clocks and quantum simulators.

\begin{figure}[ht]
\centering
  \includegraphics[width=1.0\columnwidth]{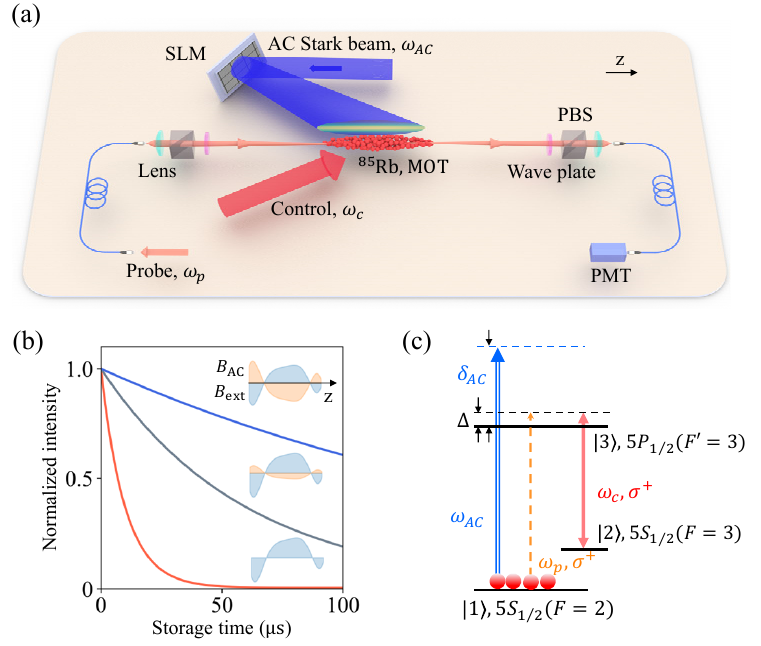}
\caption{Experimental scheme for controlling atomic optical storage with fictitious magnetic fields. (a) Optical setup. A pulsed probe signal, $\omega_p$, is stored via EIT by a control beam, $\omega_c$, in an elongated atomic cloud. The retrieved signal, $\omega_p$, is detected after storage by a high-gain photomultiplier (PMT). An AC Stark shift laser beam, $\omega_{AC}$, is utilized to compensate for inhomogeneities that lead to storage decay. (b) Mechanism of compensation of non-uniform energy shifts by fictitious magnetic fields. When $B_{AC}=-B_{ext}$ the maximum storage is reached. (c) $\Lambda$ configuration for atomic optical storage and the AC Stark laser beam. The atoms are initially prepared in the ground state $\vert 1\rangle$. The probe and the control beams are both circularly polarized. All these laser beams have a wavelength of approximately $\SI{795}{\nano\meter}$. }
\label{Fig1}
%Figure dimensions 18.8 x 14.87 cm
% \end{figure*}
\end{figure}

\textit{Experimental setup and theory.} We use an elongated cloud of cooled rubidium atoms to perform optical storage, as shown in Fig.~\ref{Fig1}(a). The atoms are prepared to the lowest hyperfine manifold $\vert 1 \rangle$ in a dark-line two-dimensional magneto-optical trap (MOT)~\cite{zhang_darkline_2012}. After MOT loading, the atoms are released, and a \SI{1.7}{\milli\second} experimental window of light storage and retrieval starts. A submicrowatt pulse of laser light, $\omega_p$, is stored as atomic coherence by a control beam, $\omega_c$, utilizing an electromagnetically induced transparency (EIT) storage scheme when $\Delta = 0$~\cite{EIT2005} or a Raman one for $\Delta \neq 0$~\cite{reim2010towards}. An AC Stark shift beam, $\omega_{AC}$, is spatially engineered to have a highly controllable intensity or polarization profile by a spatial light modulator~\cite{PhysRevApplied.15.054034, Moreno2012}, while its temporal waveform can also be synthesized by arbitrary signal generators. In general, the complex Rabi frequency of this AC Stark shift beam, which is used to compensate for unwanted Zeeman shifts and thus extend the lifetime of the stored atomic coherence, can take the spatiotemporally modulated form of $\Omega_{AC}\left(x, y, z, t\right)$. In Fig.~\ref{Fig1}(b), we elucidate the physical mechanism that enables us to rectify spatial inhomogeneities (or temporal fluctuations if $z\rightarrow t$) of the spin wave distribution. The three-level system for light storage in $^{85}$Rb together with the AC Stark beam is shown in Fig.~\ref{Fig1}(c)~\cite{Steck85_2024}.

In general, the Hamiltonian of a multi-level atomic gas interacting with a laser beam will have spin-dependent terms. The energy shifts of the various $m_F$ states can be classified as scalar, vector, and tensor, and the spin-dependent vector shifts are the ones utilized here for controlling the storage lifetimes ~\cite{Hu2018Observation}. A single beam with non-uniform spatial profile and arbitrary polarization can generate a space dependent synthetic magnetic field. The situation is quite simpler for an AC Stark shift beam interacting with $^{85}$Rb atoms with a detuning $\delta_{AC}$ larger than the hyperfine splitting ($\Delta E_{hfs} =\SI{3.036}{\giga\hertz}$), and at the same time much smaller than the fine structure splitting ($\Delta E_{fs} =\SI{7.1}{\tera\hertz}$)~\cite{Steck85_2024}. The contribution of the tensor term is negligible, while the scalar term causes $m_F-$independent shifts, which are at the same time much smaller than the spectral window of the EIT, so they can be ignored for space independent fictitious fields~\cite{SM_2024}. In this case, the experimentally relevant part of the Hamiltonian $H$ will be approximated only by a spin-dependent, Zeeman-like vector term $H_v$ that for a given detuning $\delta_{AC}$ is proportional to the intensity of the AC beam, as well as the involved $m_F$ state~\cite{SM_2024}. This vector term takes the simple form:
\begin{equation}\label{Hv}
    H_v\left(x, y, z, t\right)  =  q\mu_{AC}m_F|\Omega_{AC}\left(x, y, z, t\right)|^2,
\end{equation}
\noindent where $q={0,\pm1}$ represents respectively linearly and circularly polarized AC light, $\mu_{AC}$ depends on the associated atomic transitions and can be considered an effective Bohr magneton-like factor that converts fictitious magnetic fields to energy shifts. This fictitious magnetic field term $H_v$ can in principle be spatiotemporally modulated following the Rabi frequency $\Omega_{AC}$~\cite{Leszczynski2018Spatially}.

Equation~\ref{eta} provides a direct way to elucidate the physical mechanism that relates magnetic fields and the time dependence of quantum storage retrieval efficiency $\eta$~\cite{Peters2009, Zhao2009Longlived, Choi2011Coherent, Veissier2013Quantum, SM_2024}. 

\begin{widetext}
\begin{equation}\label{eta}
\eta\left(t\right) = |\langle \Psi_{t} | \Psi_{t=0} \rangle|^2 = |\sum_{m_F,m_{F'}}{a_{m_F} \exp{\left[i\mu_B g_{FF'}\left(m_F,m_{F'}\right)\left(B_{0}+B_{1} z+B_{2}z^2\right)t\right]}}|^2\exp\left(-\Gamma t\right).
\end{equation}
\end{widetext}

\noindent In Eq.~\ref{eta}, for notation simplicity we do not differentiate the standard from the light induced magnetic fields. $a_{m_F}$ are the factors associated with the population of the $m_F$ states and the strength of the relevant two-photon transitions for storage. $g_{FF'}=m_{F'}g_{F'}-m_{F}g_{F}$, where $g_F$ and $g_{F'}$ are the g-factor of the corresponding atomic levels. For same polarization $q$ for the probe and control beams of the storage scheme $m_{F'}=m_{F}+q$. The magnetic field terms $B_{0,1,2}$ refer to bias, gradient, and curvature, respectively. The exponential decay factor can also be replaced by a Gaussian one depending on the origin of any additional decoherence mechanisms that do not stem from magnetic fields, such as motional dephasing~\cite{Zhao2009Millisecond}.
%\section{Controlling storage by AC Stark shifts}
\begin{figure}
\centering
  \includegraphics[width=1.0\columnwidth]{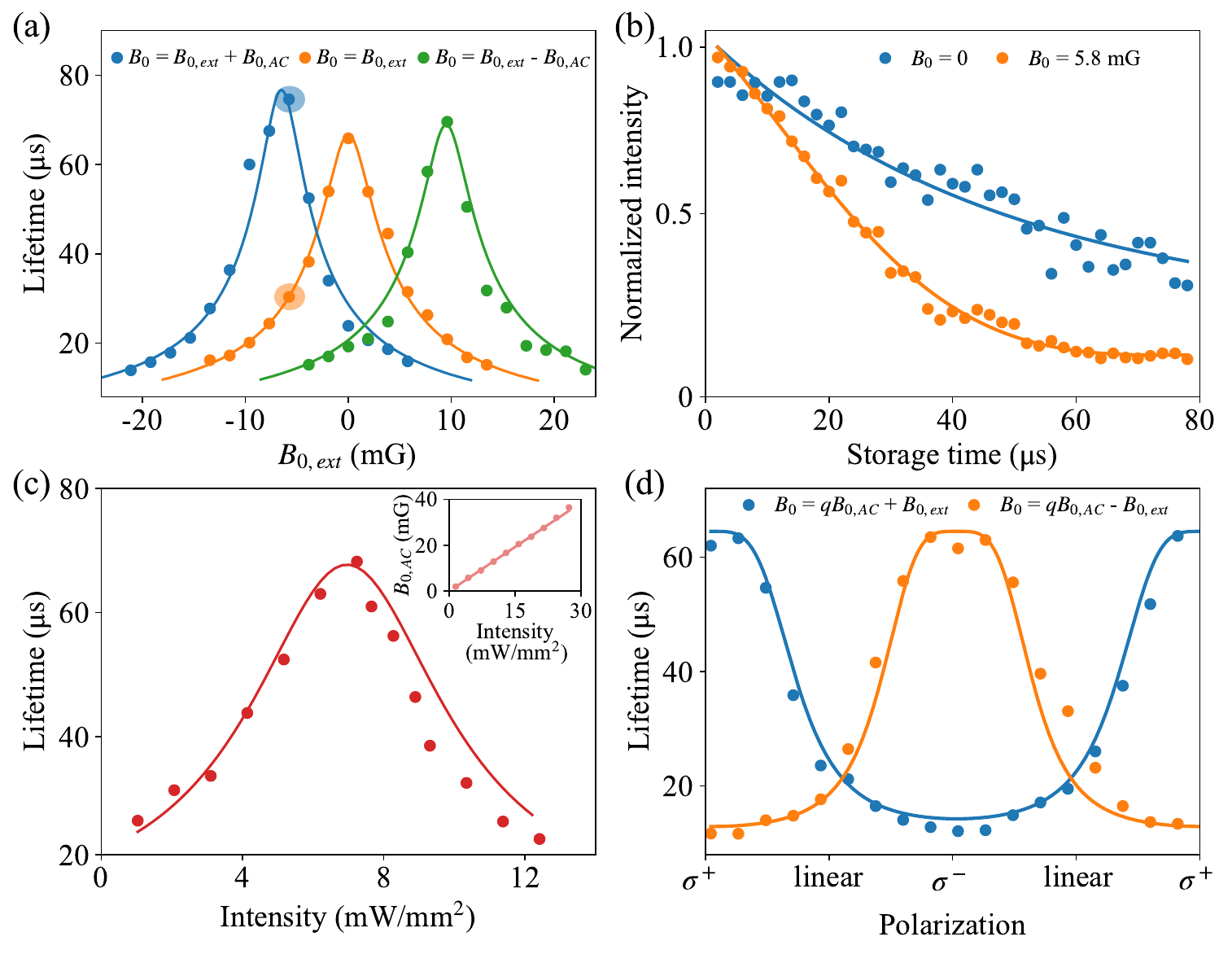}
\caption{Controlling optical storage in an atomic ensemble with an AC Stark beam. (a) Storage lifetime as a function of compensation coil magnetic field along the direction $z$. The three curves correspond to AC Stark shifts equivalent to $\SI{6.7}{\milli G}$ (blue), $\SI{0}{\milli G}$ (orange), and $\SI{-9.6}{\milli G}$ (green). (b) The orange curve illustrates the relationship between storage efficiency and storage time at a residual magnetic field of $\SI{5.8}{\milli G}$. The blue curve represents the result compensated by fictitious magnetic field. These curves correspond the shadowed discs of (a). (c) Storage lifetime as a function of the AC Stark beam intensity. The inset shows the linear dependence of the fictitious magnetic field on the intensity of an AC Stark beam. (d) Storage lifetime as function of the polarization of the AC Stark shift beam for two different initial external magnetic fields $\SI{9.6}{\milli G}$ (orange) and $\SI{-7.7}{\milli G}$ (blue). The reported fields are the ones experimentally determined, while the fitting curves ~\cite{SM_2024} correspond to theoretical fields scaled by $0.5$.}
\label{Fig2}
%Figure dimensions 18.8 x 14.87 cm
\end{figure}

\textit{Controlling storage by AC Stark shifts.} Storage of light in a cold atomic ensemble is particularly sensitive to spin-dependent energy shifts that commonly originate from uncompensated magnetic fields~\cite{Peters2009}. In Fig.~\ref{Fig2}, we experimentally demonstrate how an AC Stark beam that emulates synthetic magnetic fields can optically control the storage in an atomic memory. Here, to create a spatially uniform fictitious magnetic field along the long axis of the MOT we use an AC beam with a waist of $\SI{1}{\milli\meter}$ propagating practically parallel to this axis ($\theta=\SI{3}{\degree}$) with a detuning $\delta_{AC}=2\pi \times \SI{25.6}{\giga\hertz}$. 

In Fig.~\ref{Fig2}(a), we show that the energy shifts due to the AC beams create effective bias magnetic fields that with respect to storage lifetimes are indistinguishable from fields generated from standard magnetic coils~\cite{cohen1972experimental}. To demonstrate this, the current in a compensation coil along the direction $z$ is varied when no AC beam is present or when an AC beam, with an intensity of $I_{AC}=\SI{4.5}{\milli\watt/\milli\milli^2}$ and circular polarization $\sigma^+$ or $\sigma^-$ is applied. The reversal of polarization, as indicated in Eq.~\ref{Hv}, leads to opposite magnetic fields, while the distance of the peaks is a direct measure of the magnetic field amplitudes~\cite{cohen1972experimental}. 

Two typical storage decay curves are presented in Fig.~\ref{Fig2}(b). When the residual bias magnetic field along the MOT axis is uncompensated a shorter lifetime is observed~\cite{Peters2009}, when the AC Stark beam creates a bias-like energy shift the storage is restored. The AC induced field is experimentally estimated to be $\SI{5.8}{\milli G}$ and has a direction along $z$. We attribute a nominal field zero for the maximum lifetime achieved only by bias compensation. 

We illustrate the effect of the AC Stark beam optical power to the light storage in Fig.~\ref{Fig2}(c). Each data point is retrieved from the $1/e$ decay of curves similar to Fig.~\ref{Fig2}(b), while the fitting curves are calculated based on Eq.~\ref{eta}. A slight asymmetry on the high-power side can be attributed to optical pumping and scattering due to the AC beam~\cite{SM_2024}. Our assumption is that these effects pose the main limits on the maximum effective magnetic fields reachable.  Fig.~\ref{Fig2}(c) (inset) shows the expected linear dependence of the AC induced magnetic fields on the optical intensity. 

The polarization of an AC Stark shift can provide an additional degree of freedom and be used to generate arbitrary magnetic fields~\cite{cohen1972experimental}. In Fig.~\ref{Fig2}(d), the storage lifetime is measured when the polarization is tuned continuously from left to right circular passing from the zero-magnetic fields linear case. %This is demonstrated starting from two different bias field values $\SI{9.6}{\milli G}$ and $\SI{-7.7}{\milli G}$.

\begin{figure}
\centering

  \includegraphics[width=1.0\columnwidth]{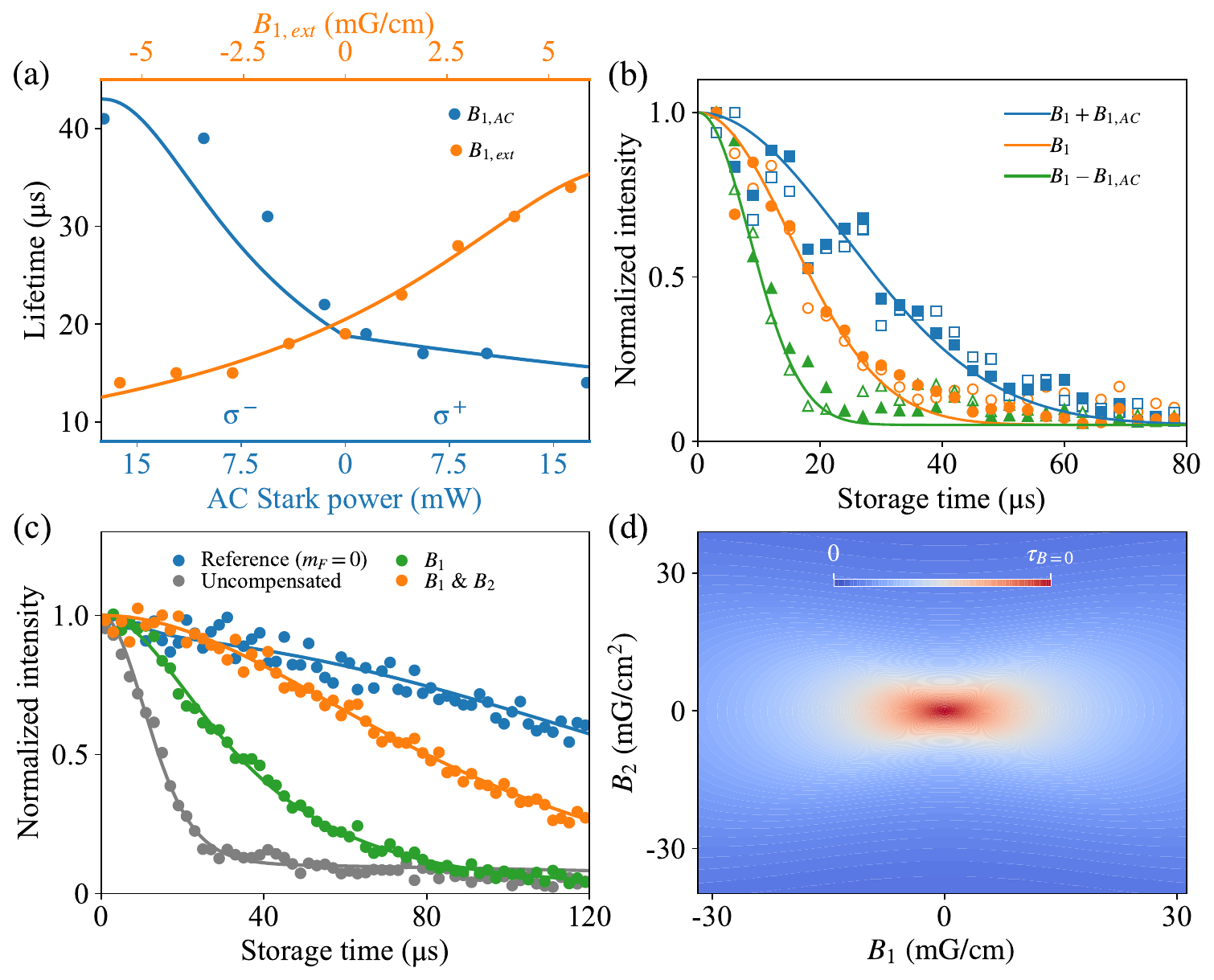}
\caption{Compensating spatial inhomogeneities with AC Stark beams. (a) Storage lifetime dependence on the gradient of linearly dependent effective magnetic fields (blue line) induced by an AC Stark beam for $\sigma^+/\sigma^-$ polarizations, and standard magnetic fields with opposite gradient $B_{1,ext}$ (orange line). The lifetimes are the $1/e$ points for curves similar to (b). (b) Retrieved pulse intensities as a function of time for optically induced gradient, $\pm B_{1,AC}$ (green/blue open markers), and standard magnetic field gradient $\pm B_{1,ext}$ (green/blue solid markers). The orange curve and data correspond to the storage decay when no compensation is applied (orange open markers) and when the two fields opposite to each other (orange solid markers). (c) Storage lifetime dependence for the state $m_F = 1$ when no compensation fictitious magnetic fields are applied ($B_{AC}=0$, gray line), when a gradient field partially compensates the nonuniformities due to a quantization field ($B_{AC}=+B_{1,AC} z$, green line), and when a quadratic term is also added ($B_{AC}=B_{1,ext} z+B_{2,AC} z^2$, orange line). The reference (blue line) is the storage decay starting from the non-magnetic $m_F=0$ state. (d) Theoretical prediction of the storage lifetimes as a function of uncompensated linear $B_1$ and quadratic terms $B_2$ for the $m_F=1$ state. The colormap is normalized to a maximum storage time $\tau_{B=0}$ that is independent of magnetic fields.}
\label{Fig3}
%Figure dimensions 18.8 x 14.87 cm
\end{figure}

\begin{figure*}

  \includegraphics[width=2.0\columnwidth]{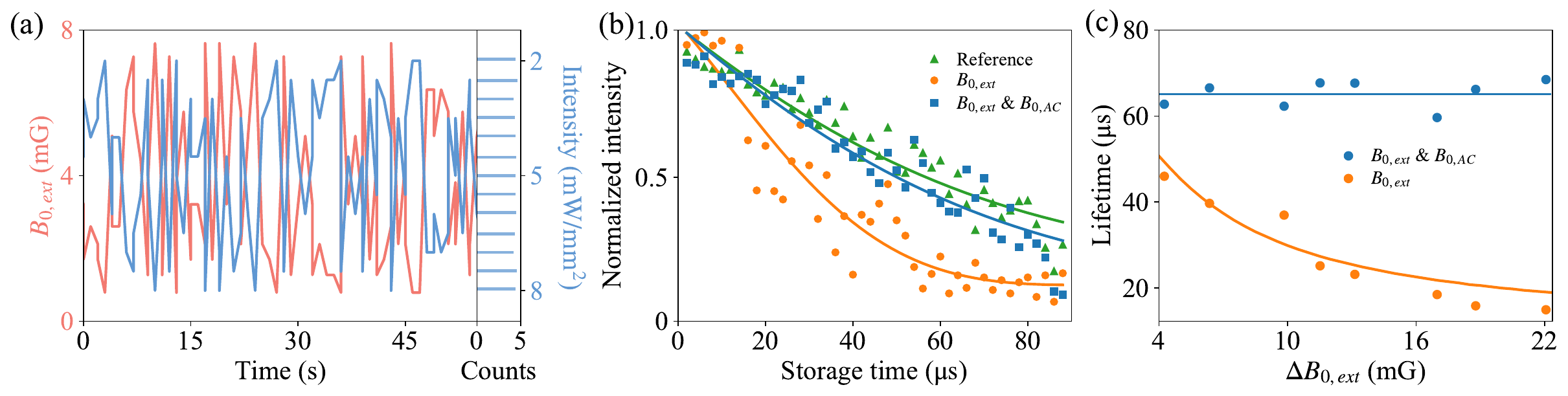}
\caption{Controlling optical storage in an atomic ensemble with a time-varying AC Stark beam. 
(a) (Left) Time series of the random bias magnetic field (red line) changing in 1Hz, and the corresponding compensating fictitious field (blue line).
(Right) a histogram of the uniform intensity distribution of the $N=13$ values of the AC stark beam. (b) Retrieved pulse intensity as a function of storage time when only a time-varying control magnetic field in a range of $\Delta B_{0,ext} = \SI{6.8}{\milli G}$ (orange circles) is applied, when a compensating AC beam is applied simultaneously with the external magnetic field (blue squares), and when no perturbing magnetic or optical fields are applied (green triangles). 
The solid lines represent the corresponding theoretical fits. 
(c) Storage lifetimes as a function of time-varied applied magnetic fields in the range $\left\langle \Delta B_{0,ext} \right\rangle$ are acting solely (orange circles), and for the case that the AC beam is compensating the field variations (blue circles). The solid lines are a constant (blue line) and a fit from statistical averaging of the fluctuating magnetic fields (orange line).}
\label{Fig4}
%Figure dimensions 18.8 x 14.87 cm
\end{figure*}

%\section{Space-dependent fictitious magnetic fields}

\textit{Space-dependent fictitious magnetic fields.} Precise manipulation of fictitious magnetic fields in the spatial domain can be instrumental in correcting for inhomogeneities that stem from non-uniform stray magnetic fields. Recent experiments on quantum storage of photons in cold atomic ensembles have attributed to certain limitations to the storage lifetimes to inhomogeneities in the range of $\Delta B = \SI{10}{\milli G}$~\cite{Hsiao2018Highly, felinto2005}. In Fig.~\ref{Fig3}, we demonstrate that AC Stark shifts can introduce effective magnetic fields that match the most commonly occurring lowest-order inhomogeneities such as linear and quadratic. We have utilized a Raman storage scheme~\cite{reim2010towards,guo2019high} that addresses only one of the available ground magnetic states $m_F$, and therefore the contributions of the bias fields are decoupled from the storage lifetimes. Spatial modulation of the AC Stark shift beam has been done by a spatial light modulator (SLM)~\cite{PhysRevApplied.15.054034, Moreno2012} as shown in Fig.~\ref{Fig1}(a) and is detailed in~\cite{SM_2024}.

%%%First, standard magnetic field gradients have also been used to illustrate their equivalence with the fictitious Stark induced fields. 

As shown in Fig.~\ref{Fig3}(a), with double axes we demonstrate the similar dependence of storage on magnetic field gradients on a current coil and the AC-Stark beam. The fitting curves are based on Eq.~\ref{eta}, for gradients with a maximum value of $\SI{9}{\milli G/\centi\meter}$ and $\SI{2}{mG/cm}$ for polarizations $\sigma^-$ and $\sigma^+$ respectively, with optical power $P = \SI{17.4}{\milli\watt}$ for a red detuning of $\SI{17.2}{\giga\hertz}$. In Fig.~\ref{Fig3}(b), we demonstrate that optically induced AC Stark shifts with linear space dependence can correct for the weak residual gradient magnetic fields existing in the environment. This residual field leads to a storage lifetime of approximately $\SI{22}{\micro\second}$ (orange line) for the state $m_F=1$. When the AC Stark beam with linear dependence and positive slope, due to the polarization $\sigma^-$, is applied, $+B_{1,AC} z$, the lifetime can be improved (blue line). On the other hand, if the polarization is reversed the effective field gradient will be opposite, $-B_{1,AC} z$, and the storage will be severely reduced (green line).  We have also demonstrated that an AC induced gradient (solid markers) can have identical effect with a purely current induced magnetic gradient (open markers) and cancel external magnetic field gradients in the order of several $\mathrm{mG/cm}$.

The highest degree of cancellation of magnetic fields would lead to the stored light in magnetic states reaching the magnetically insensitive state $m_F=0$ that acts as a reference (blue line). With that target, we have compensated the linear gradient (green line), and we have added a parabolically modulated AC Stark beam $B_{2,AC}z^2$ (orange line) and further improvement of storage lifetime of the completely uncompensated decay (gray line) as shown in Fig.~\ref{Fig3}(c). By this approach, we have achieved more than a five-fold improvement of the storage lifetime: from $\SI{18}{\micro\second}$ (gray line) to $\SI{100}{\micro\second}$ (orange line). The remaining gap between the highest $m_F=1$ lifetime and the $m_F=0$ can be attributed to higher terms of the uncompensated magnetic fields and contributions from the stray RF fields. The lifetime of the reference state itself is currently predominantly limited by dephasing stemming from thermal motion ~\cite{Zhao2009Millisecond, SM_2024}. In Fig.~\ref{Fig3}(d), we complement our experimental observations with theoretical predictions for the combined effect of linear $B_1$ and quadratic terms $B_2$ on the lifetime of light storage in a cold atomic ensemble of $m_F=1$ atoms~\cite{Choi2011Coherent,Veissier2013Quantum}.

%\section{Time-dependent fictitious magnetic fields}

\textit{Time-dependent fictitious magnetic fields.} While compensation of stray magnetic fields is routinely performed with standard current carrying coils, AC Stark shifts  due to their optical origin have two distinct features that make them compelling for application in quantum storage: they can be spatially sculpted in the micrometer scale in arbitrary patterns, and  in the temporal domain can be readily modulated even in the nanosecond regime. A proof-of-concept demonstration of temporal compensation of external magnetic fields is presented in Fig.~\ref{Fig4}. Magnetic field variations can be detrimental for an atomic memory and significantly reduce the quality of photon storage via minuscule random fluctuations during the storage time or via larger longer-term instabilities.  Here, in Fig.~\ref{Fig4}, we demonstrate a method to address the longer-term instability issue by compensating a random external magnetic field by employing an AC beam that induces an opposite effective magnetic field. The left panel of Fig.~\ref{Fig4}(a) shows the time series of the two opposing fields, while its right panel shows a histogram of the $N=13$ intensity levels of the AC Stark beam (blue line) applied to compensate for the corresponding magnetic fields (red line). %More concretely, an atomic ensemble in the lowest hyperfine manifold of $^{85}$Rb is prepared in a MOT every $\SI{18.3}{\milli\second}$ and then it is released to perform storage. 
The random, Fig.~\ref{Fig4}(a), but known control magnetic field that can be varied in every experimental cycle is applied to the atoms and the atomic storage decay is observed. As seen in Fig.~\ref{Fig4}(b), a reduced storage lifetime is observed when no compensation beams are used (orange line). We observed a similar behavior when a calibrated AC beam was applied in isolation. When the control magnetic field and the AC beam are applied together the storage lifetime is restored (blue line) to the no magnetic field case (green line).

To demonstrate the robustness of this technique, we have repeated, as shown in Fig.~\ref{Fig4}(c), this experiment for different peak-to-peak random modulations of the magnetic field and were able to compensate them and restore the storage lifetime to its initial value of approximately $\SI{65}{\micro\second}$ (blue line). The storage lifetimes without AC compensation naturally follow an exponential-like decay (orange line). This decay is in agreement with numerical simulations that average the storage decays for fluctuating magnetic fields with increasing peak-to-peak variations, but with a scaling factor of approximately $0.5$.

%\section{Conclusions}
\textit{Conclusions.} In this work, we have demonstrated that optically generated fictitious magnetic fields can extend the lifetime of quantum memories in cold atomic ensembles. All the degrees of freedom of an AC Stark beam such as polarization, spatial profile, and temporal waveform can be readily controlled in a precise manner, and remarkable improvements of the storage lifetime are recorded. AC Stark shift beams have the potential to create synthetic magnetic fields with sub-micrometer spatial resolution and tens of nanoseconds switching times. The aforementioned advantages of fictitious magnetic fields are particularly pronounced for effective magnetic fields up to a few tens $\SI{}{\milli G}$ and can be readily used to address inhomogeneities corresponding even down to $\SI{}{\micro G}$. A limitation of the demonstrated method is that inhomogeneities of the scalar terms of the AC Stark shifts themselves cannot be simultaneously compensated for all $m_F$ states. An immediate straightforward next step would be to demonstrate inhomogeneity compensation in the spatiotemporal domain by machine learning~\cite{Tranter2018Multiparameter}.  While we here focus on perturbations stemming from magnetic fields our method can find application in other disturbances originating for instance on imperfect profiles of the write and read beams. The versatility of the AC induced fictitious fields can complement standard magnetic fields for any storage scheme for other atomic systems, and we envision that can be exploited for realizing periodic artificial gauge fields to manipulate dark state polaritons that are featuring a hybrid atom-photon wavefunction~\cite{Lukin2000, hammerer2010}. Space dependence can be used to engineer the momentum of the stored spin waves, and the temporal periodicity can lead to Floquet engineered effective Hamiltonians similar to  the ones emerging for photons~\cite{Ozawa_2019} and quantum degenerate gases~\cite{cooper_topological_2019}.

%\section{Acknowledgments}

\begin{acknowledgments}
\textit{Acknowledgments.} This work is supported by the National Natural Science Foundation of China (NSFC) through Grants No. 12074171, No. 12074168, No. 92265109, and No.12204227; the Guangdong Provincial Key Laboratory (Grant No. 2019B121203002), and the Guangdong projects under Grant No. 2022B1515020096 and No. 2019ZT08X324. X. W. acknowledges the support from the SUSTech Presidential Postdoctoral Fellowship. 
\end{acknowledgments}

\bibliography{main} 

\pagebreak 

\clearpage

\widetext

\appendix

\renewcommand{\appendixname}{}
% \renewcommand{\thefigure}{S\arabic{figure}}
% \setcounter{figure}{0}
% \setcounter{equation}{0}

% \counterwithout{equation}{section}
% \renewcommand{\theequation}{S\arabic{equation}}

\begin{center}
    {\Large \textbf{Supplemental Material for ``Light-induced fictitious magnetic fields for quantum storage in cold atomic ensembles"}}
\end{center}

\maketitle

%\clearpage

\section{Cold atoms experimental setup}\label{sec01}

$^{85}$Rb atoms are prepared in a two-dimensional dark-line magneto-optical trap (MOT) with waists of \SI{2.5}{\centi\meter} and \SI{500}{\micro\meter} in the longitudinal and transversal directions, respectively. Three orthogonal pairs of counter-propagating laser beams, with frequency of $\SI{20}{\mega\Hz}$ red detuned from the transition $ \vert 5^2S_{1/2}, F=3 \rangle \rightarrow \vert 5^2P_{3/2},F'=4\rangle$ together with two counter-propagating repumping laser beams, with frequency of $\vert 5^2S_{1/2}, F=2\rangle \rightarrow \vert 5^2P_{3/2}, F'=2 \rangle$ cool the atoms to around \SI{120}{\micro\kelvin}. These beams have a waist of $\SI{18}{\milli\meter}$ and they intersect on the zero-field line parallel to the long axis of a rectangular coil that creates a magnetic field gradient of $\SI{7}{G/ \cm}$. The optical depth (OD) of the MOT is around $110$ in our storage experiment.

\section{Time sequence for the atom preparation and OD characterization}\label{sec02}

\begin{figure*}[htbp]
  \centering
  \includegraphics[width=0.7\linewidth]{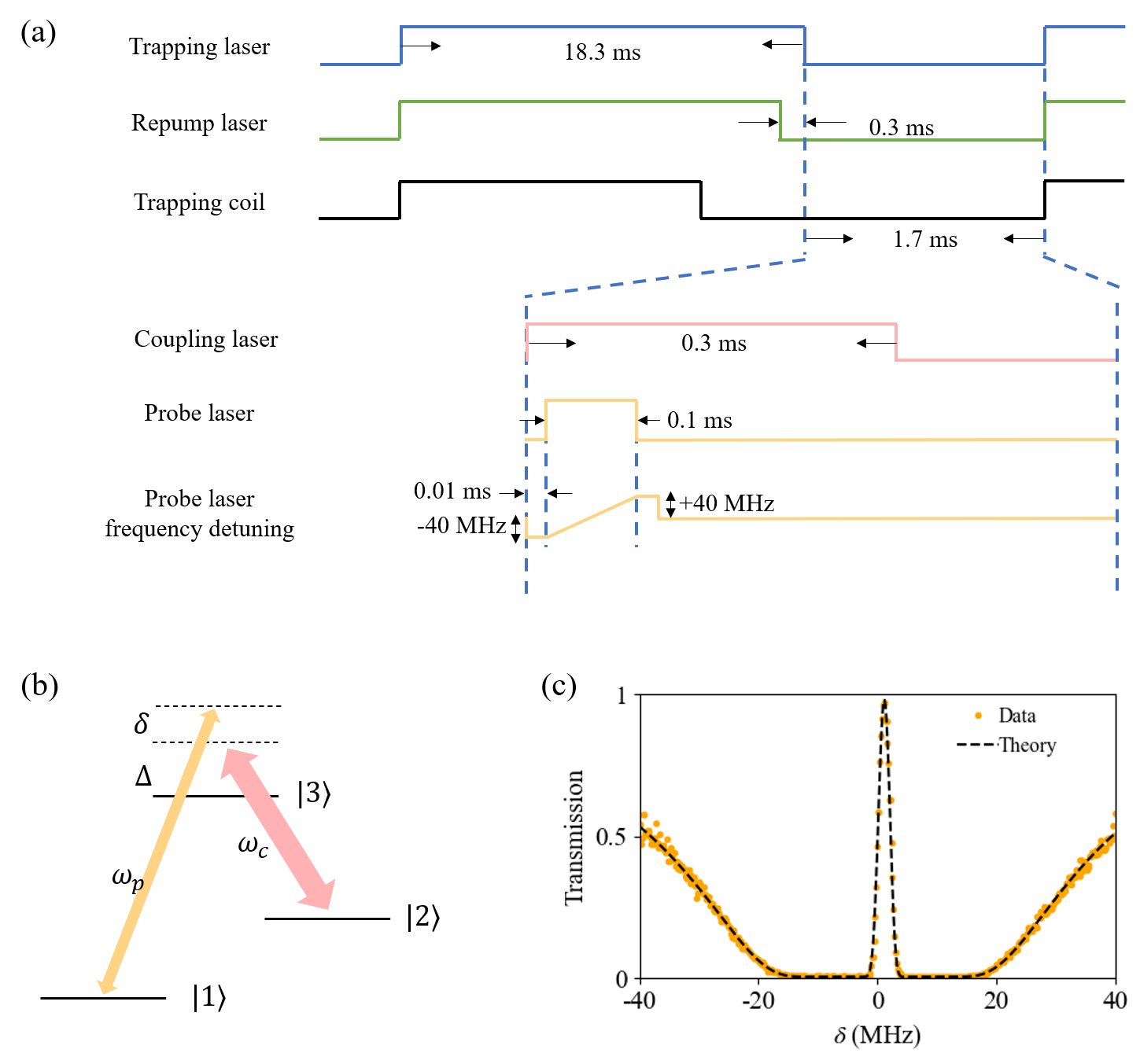}
\caption{MOT preparation and optical depth characterization (a) The time sequence. (b) Energy levels for the coupling and probe lasers, where $|1\rangle = |5 S_{1/2},F=2\rangle $, $|2\rangle = |5 S_{1/2},F=3\rangle $ and $|3\rangle = |5 P_{1/2},F'=3\rangle $. $\Delta$ is the detuning of the coupling laser with respect to the $|2\rangle \to |3\rangle$ transition, and $\delta$ refers to the detuning of two-photon transition used for EIT storage. (c) Spectrum of electromagnetically induced transparency.}
\label{Fig_SM_A1}
\end{figure*}

To efficiently store light pulses, the initial MOT preparation and the storage measurements follow the time sequence shown in Fig.~\ref{Fig_SM_A1}(a) with a repetition rate of 50Hz, where the atoms are prepared in $\SI{18.3}{\milli \second}$, and the measurement window is $\SI{1.7}{\milli \second}$.
The trapping laser is switched on during the MOT preparation time of $\SI{18.3}{\milli \second}$ while the repumping laser is switched off $\SI{0.3}{\milli \second}$ in advance for transferring the atoms to the ground state manifold $\vert1\rangle$. 
The current of the MOT magnetic coil is turned off  $\SI{1.5}{\milli \second}$ in advance to avoid unwanted magnetic fields on the storage.
During the measurement window, the OD is characterized by electromagnetically induced transparency (EIT), where the probe beam is switched on for $\SI{100}{\micro \second}$ with its frequency detuning swept linearly from $\SI{-40}{\mega\Hz}$ to $\SI{40}{\mega\Hz}$ relatively to the $\vert1\rangle\to \vert3\rangle$ transition, with the coupling laser on resonance with  $|2\rangle\to |3\rangle$ in presence as Fig.~\ref{Fig_SM_A1}(b) shows. The EIT spectrum is shown in Fig.~\ref{Fig_SM_A1}(c) and the fitted OD is 114.

\section{Time sequence for the light storage measurement}\label{sec03}

\begin{figure*}[htbp]
  \centering
  \includegraphics[width=0.7\linewidth]{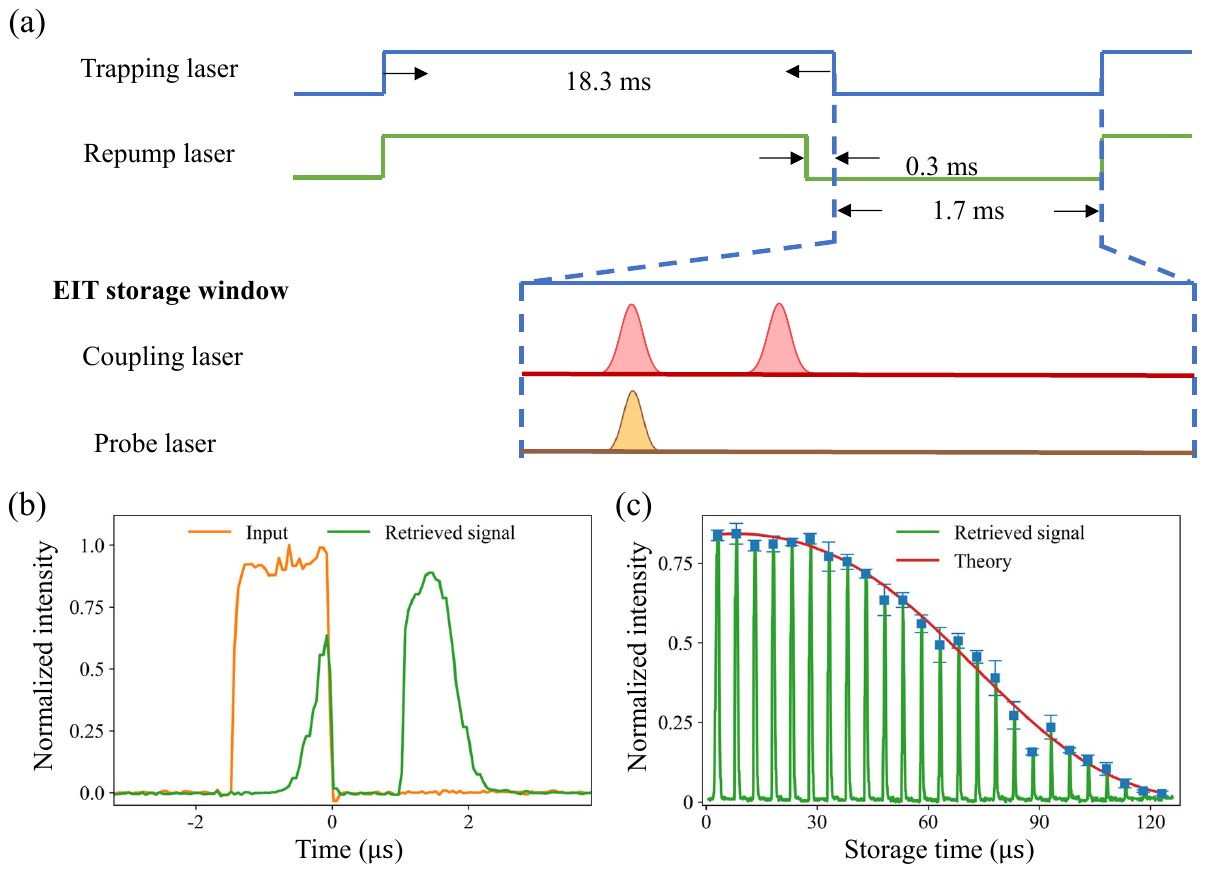}
\caption{Light storage measurement with EIT scheme and Raman scheme (a) The time sequence. (b) Signal of the input probe pulse (orange line) and retrieved probe pulse after \SI{1}{\micro\second} of storage (green line). (c) Experimentally retrieved pulses (green lines) and their peak intensity (blue point) as a function of storage time together with the theoretical fit.}
\label{Fig_SM_A2}
\end{figure*}

In the presence of the AC Stark beam, the light storage and retrieval are performed after MOT preparation as Fig.~\ref{Fig_SM_A2}(a) shows.  We apply two different storage schemes, EIT storage and Raman storage.  
For both storage schemes, a coherent probe pulse of typically less than \SI{10}{\nano\watt} and a pulse width of around \SI{1}{\micro\second} is turned on after the release of the atoms from the MOT. 
The coupling laser is turned on following the pulse shape of the probe laser to slow down the probe pulse with EIT or complete the Raman transition in two different schemes.  The stored probe pulse is retrieved by another coupling pulse applied after a certain storage time completing the writing process as Fig.~\ref{Fig_SM_A2}(b) shows. 
The AC Stark beam that induces the fictitious magnetic fields is continuously applied during the whole experiment cycle. 
In the EIT storage scheme, the coupling and probe laser are on resonance with the relevant transitions ($\Delta=0$ and $\delta=0$), and since the quantization field is not applied the Zeeman sublevels are degenerate. 
In the Raman storage scheme, the coupling and probe lasers are both far-off resonance ($\Delta = 90$MHz) from the respective atomic transitions while the two-photon detuning is $\delta = 0$. A pair of Helmholtz coils provides a quantization magnetic field along the longitudinal direction of the MOT and separates the Zeeman sublevels during the storage window.  For the narrow Raman transition width of our experiments, only one $m_F$ state of each hyperfine level is involved in storage. The energy levels are changed accordingly as $|1\rangle = |5S_{1/2}, F = 2, m_{F} = j\rangle$, $|2\rangle = |5 S_{1/2}, F=3, m_F = j\rangle$, and $|3\rangle = |5P_{1/2}, F'= 3, m_{F'} = j+1\rangle$, with both probe and coupling lasers being circularly polarized. 

We ignore the transmitted portion of the pulses that failed to be stored, and the pulses retrieved with different time are recorded and analyzed as Fig.~\ref{Fig_SM_A2}(c) shows. The storage lifetime is defined as the storage time where the retrieval efficiency is decreased to $1/e$. Typical Rabi frequencies for the coupling and the signal pulses are $\Omega_{c}=3.0\gamma_{13}$ and $\Omega_{p}=0.001\gamma_{13}$, where $\gamma_{13}=2\pi\times\SI{3}{MHz}$ denotes the dephasing rate of the $\vert3\rangle\rightarrow\vert1\rangle$ transitions. 

\section{Theory of the fictitious magnetic fields in $ ^{85}$Rb}\label{sec04}

A general theory of optically induced fictitious magnetic fields for atoms together with a first demonstration has been introduced in~\cite{cohen1972experimental}. In most of our experiments, the conventional magnetic field axis almost coincides with the propagation direction of the AC Stark beam and thus the fictitious magnetic fields will be also along the same axis. Artificial magnetic fields can be decomposed into the scalar, vector, and tensor components. For Fig.~2 and Fig.~4 of the main text, only the vector magnetic fields, which have a linear dependence on $m_F$, have influence on the storage since the scalar ones represent a common energy shift for all $m_F$ components, while the tensor ones are too small to affect the storage in the sub-millisecond timescale as is shown below. For Fig.~3 which deals with spatially inhomogeneous AC Stark shifts, we have implemented a Raman storage scheme that uses a single $m_F$ state to avoid subtleties related with the fact that spatially dependent components cannot be cancelled for all spin states simultaneously. The AC Stark beam has a wavelength of $\SI{795}{\nano\meter}$ and is typically $\delta_{AC}=10-20\SI{}{\giga\Hz}$ red or blue detuned from the $|1\rangle \to |3\rangle$ transition. 

After a light pulse has been stored only the $m_F$ states of the $|1\rangle$ and $|2\rangle$ levels are important since the excited $|3\rangle$ is common for both arms of the two-photon transition and thus its shift is canceled. The relevant AC Stark shifts of the vectorial term can be well approximated by a simple formula:
\begin{equation}\label{A1-1}
 \Delta E_{AC,F} = \frac{q m_F I\Gamma}{\delta_{AC,F}}
 \end{equation}
where $\delta_{AC,F}$ is the detuning from the relevant transitions and $\Gamma$ is their natural linewidth, $I$ is the intensity of the AC Stark beam, $q$ is its polarization, and $m_F$ is the state on the $F$ manifold~\cite{Leszczynski2018Spatially}. Below we present the differential energy shifts appearing in our storage experiments. We have followed~\cite{Hu2018Observation} and calculated all the energy shifts for $^{85}$Rb. The complete calculation is in agreement with the simplified model of Equation 1. The highest values of the fictitious magnetic fields generated in our experiments are approximately $\SI{50}{\milli G}$ for powers less than $\SI{100}{\milli\watt}$.

The energy level shift caused by light field can be calculated:
\begin{equation}\label{A1}
\Delta E^{AC}(\omega)=-\alpha_{F,m_F}(\omega)\left(\frac{E}{2}\right)^2,
\end{equation}
where $\alpha_{F,m_F}(\omega)$ is the dynamic polarizability. $E$ and $\omega$ are the amplitude and frequency of the laser field. The dynamic polarizability is usually expressed in a form with $m_F$-independent factors (scalar) and $m_F$-dependent factors (vector and tensor):
\begin{equation}\label{A2}
\alpha_{F, m_F}\left(\omega\right)=\alpha_F^S\left(\omega\right)+\left(\hat{k} \cdot \hat{B}\right) q \frac{m_F}{2 F} \alpha_F^V\left(\omega\right)+\left(3|\hat{\zeta} \cdot \hat{B}|^2-1\right) \frac{3 m_F^2-F\left(F+1\right)}{2 F\left(2 F-1\right)} \alpha_F^T\left(\omega\right),
\end{equation}
where $\hat{B}$ and $\hat{k}$ are the unit vector of the quantization magnetic field and the laser field, respectively. $\hat{\zeta}$ is the complex polarization vector of the laser.
\begin{equation}\label{A3}
\begin{aligned}
&\begin{gathered}
\alpha_F^S(\omega)=\sum_{F^{\prime}} \frac{2 \omega_{F^{\prime} F}\left|\left\langle F\|\mathbf{d}\| F^{\prime}\right\rangle\right|^2}{3 \hbar\left(\omega_{F^{\prime} F}^2-\omega^2\right)} \\
\alpha_F^V(\omega)=\sum_{F^{\prime}}(-1)^{F+F^{\prime}+1} \sqrt{\frac{6 F(2 F+1)}{F+1}}\left\{\begin{array}{ccc}
1 & 1 & 1 \\
F & F & F^{\prime}
\end{array}\right\} \frac{\omega_{F^{\prime} F}\left|\left\langle F\|\mathbf{d}\| F^{\prime}\right\rangle\right|^2}{\hbar\left(\omega_{F^{\prime} F}^2-\omega^2\right)}
\end{gathered}\\
&\alpha_F^T(\omega)=\sum_{F^{\prime}}(-1)^{F+F^{\prime}} \sqrt{\frac{40 F(2 F+1)(2 F-1)}{3(F+1)(2 F+3)}}\left\{\begin{array}{ccc}
1 & 1 & 2 \\
F & F & F^{\prime}
\end{array}\right\} \frac{\omega_{F^{\prime} F}\left|\left\langle F\|\mathbf{d}\| F^{\prime}\right\rangle\right|^2}{\hbar\left(\omega_{F^{\prime} F}^2-\omega^2\right)},
\end{aligned}
\end{equation}
where $\left\langle F\|\mathbf{d}\| F^{\prime}\right\rangle$ can be written as:
\begin{equation}\label{A4}
\begin{aligned}
\left\langle F\|\mathbf{d}\| F^{\prime}\right\rangle & \equiv\left\langle J I F\|\mathbf{d}\| J^{\prime} I^{\prime} F^{\prime}\right\rangle \\
& =\left\langle J\| \mathbf{d}\| J^{\prime}\right\rangle(-1)^{F^{\prime}+J+1+I} \sqrt{\left(2 F^{\prime}+1\right)(2 J+1)}\left\{\begin{array}{ccc}
J & J^{\prime} & 1 \\
F^{\prime} & F & I
\end{array}\right\}.
\end{aligned}
\end{equation}

The energy level shifts caused by the AC Stark effect induced by the light field can be decomposed into scalar, vector, and tensor parts:
\begin{equation}\label{A5}
\Delta E^{AC}\left(\omega_{AC}\right)=\Delta E^S\left(\omega_{AC}\right)+\Delta E^V\left(\omega_{AC}\right)+\Delta E^T\left(\omega_{AC}\right)
\end{equation}

\begin{equation}\label{A6}
\begin{gathered}
\Delta E_F^S\left(\omega_{AC}\right)=-\left(\frac{\varepsilon}{2}\right)^2 \alpha_F^S\left(\omega_{AC}\right) \\
\Delta E_F^V\left(\omega_{AC}\right)=-\left(\frac{\varepsilon}{2}\right)^2(\hat{k} \cdot \hat{B}) q \frac{m_F}{2F} \alpha_F^V\left(\omega_{AC}\right) \\
\Delta E_F^T\left(\omega_{AC}\right)=-\left(\frac{\varepsilon}{2}\right)^2\left(3|\hat{\zeta} \cdot \hat{B}|^2-1\right) \frac{3 m_F^2-F\left(F+1\right)}{2 F\left(2 F-1\right)} \alpha_F^T\left(\omega_{AC}\right).
\end{gathered}
\end{equation}

For the energy levels $|1\rangle$ and $|2\rangle$ that are used for our storage schemes:

\begin{equation}\label{A7}
\begin{aligned}
& \alpha_2^S(\omega)=\left[\frac{0.148 \cdot \omega_{22}}{\left(\omega_{22}^2-\omega^2\right)}+\frac{0.518 \cdot  \omega_{23}}{\left(\omega_{23}^2-\omega^2\right)}\right] \frac{d^2}{\hbar} \\
& \alpha_3^S(\omega)=\left[\frac{0.37 \cdot \omega_{32}}{\left(\omega_{32}^2-\omega^2\right)}+\frac{0.296 \cdot \omega_{33}}{\left(\omega_{33}^2-\omega^2\right)}\right] \frac{d^2}{\hbar}\\
& \alpha_2^V(\omega)=\left[-\frac{0.074 \cdot \omega_{22}}{\left(\omega_{22}^2-\omega^2\right)}+\frac{0.519 \cdot  \omega_{23}}{\left(\omega_{23}^2-\omega^2\right)}\right] \frac{d^2}{\hbar} \\
& \alpha_3^V(\omega)=\left[-\frac{0.556 \cdot \omega_{32}}{\left(\omega_{32}^2-\omega^2\right)}-\frac{0.111 \cdot \omega_{33}}{\left(\omega_{33}^2-\omega^2\right)}\right] \frac{d^2}{\hbar}\\
& \alpha_2^T(\omega)=\left[\frac{0.175 \cdot \omega_{22}}{\left(\omega_{22}^2-\omega^2\right)}-\frac{0.175 \cdot  \omega_{23}}{\left(\omega_{23}^2-\omega^2\right)}\right] \frac{d^2}{\hbar} \\
& \alpha_3^T(\omega)=\left[-\frac{0.454 \cdot \omega_{32}}{\left(\omega_{32}^2-\omega^2\right)}+\frac{0.454 \cdot \omega_{33}}{\left(\omega_{33}^2-\omega^2\right)}\right] \frac{d^2}{\hbar},
\end{aligned}
\end{equation}
where $d=\left|\left\langle J=1 / 2\|e \boldsymbol{r}\| J^{\prime}=1 / 2\right\rangle\right|$. When $\delta_{AC}= 2 \pi \times 25.6~\mathrm{GHz}$:

\begin{equation}\label{A8}
\begin{aligned}
\alpha_2^S  & = -0.1904~h \cdot \mathrm{kHz} /(\mathrm{V} / \mathrm{cm})^2\\
\alpha_3^S  & = -0.1534~h \cdot \mathrm{kHz} /(\mathrm{V} / \mathrm{cm})^2\\
\alpha_2^V  & = -0.1276~h \cdot \mathrm{kHz} /(\mathrm{V} / \mathrm{cm})^2\\
\alpha_3^V  & = 0.1529~h \cdot \mathrm{kHz} /(\mathrm{V} / \mathrm{cm})^2\\
\alpha_2^T  & = 0.0007~h \cdot \mathrm{kHz} /(\mathrm{V} / \mathrm{cm})^2\\
\alpha_3^T  & = -0.0012~h \cdot \mathrm{kHz} /(\mathrm{V} / \mathrm{cm})^2
\end{aligned}
\end{equation}

When the intensity of AC beam is $3.0 \mathrm{~mW} / \mathrm{mm}^2$ the energy shifts for $|1\rangle$ are:
\begin{equation}\label{A9}
\begin{gathered}
\Delta \omega_2^S= 2 \pi \times\SI{10.75}{\kilo\hertz}\\
\Delta \omega_2^V=2 \pi \times q \frac{m_F}{4} \SI{7.2}{\kilo\hertz}  \\
\Delta \omega_2^T= -2 \pi \times \frac{3 m_F^2-6}{12} \SI{0.04}{\kilo\hertz}
\end{gathered}
\end{equation}

When the intensity of AC beam is $3.0 \mathrm{~mW} / \mathrm{mm}^2$ the energy shifts for $|2\rangle$ are:
\begin{equation}\label{A10}
\begin{gathered}
\Delta \omega_3^S= 2 \pi \times\SI{8.66}{\kilo\hertz}\\
\Delta \omega_3^V=- 2 \pi \times q \frac{m_F}{6} \SI{2.89}{\kilo\hertz}  \\
\Delta \omega_3^T= -2 \pi \times \frac{3 m_F^2-12}{30} \SI{0.07}{\kilo\hertz}
\end{gathered}
\end{equation}

And from these formulas we have calculated the induced AC Stark shifts associated with our storage.

\section{Shaping the AC Stark shift beam with a spatial light modulator.}\label{sec05}

To precisely control the spatial profile of the AC Stark shift beam, we use a phase-only spatial light modulator (SLM, HOLOEYE PLUTO-2) in a $4f-$configuration. %The setup to generate self-defined curvature is illustrated in Fig.~\ref{Fig_SM_3}(a). 
A linearly polarized Gaussian beam with a diameter of $\SI{2}{\centi \meter}$ is incident on the SLM. The phase modulation on SLM follows the equation:
\begin{equation}
    \phi(z',y') = m(z') \sin(\frac{2\pi}{T_z'} z')
\end{equation}
where $z'$ and $y'$ are according to the axes of the longitudinal and vertical directions of the SLM, $m(z')$ is the modulation depth, and $T_z'$ is the period of the sine modulation which determines the distance between the diffraction beams in the aperture plane. The reflected light is focused to several beam spots with a $\SI{150}{\milli \meter}$ lens, and the intensity of the zero-order beam is controlled by the modulation depth $m$. We choose the $m(z')$ from $0$ to $\pi$, and the zero-order beam intensity decreases with $m(z')$ increasing. We numerically estimate the relation between the intensity of the zero-order beam and $m(z')$ as the first attempt and optimize the $m(z')$ by measuring the generated intensity distribution and then we update the $m(z')$.
On the Fourier plane, the high-order beams are filtered out with an aperture, and the zero-order beams are collimated again with a $\SI{200}{\milli \meter}$ lens. The modulated beams are incident on the MOT with an angle of $5^{\circ}$ with respect to the longitudinal direction of the MOT. 

\begin{figure*}[htbp]
  \centering
  \includegraphics[width=0.8\linewidth]{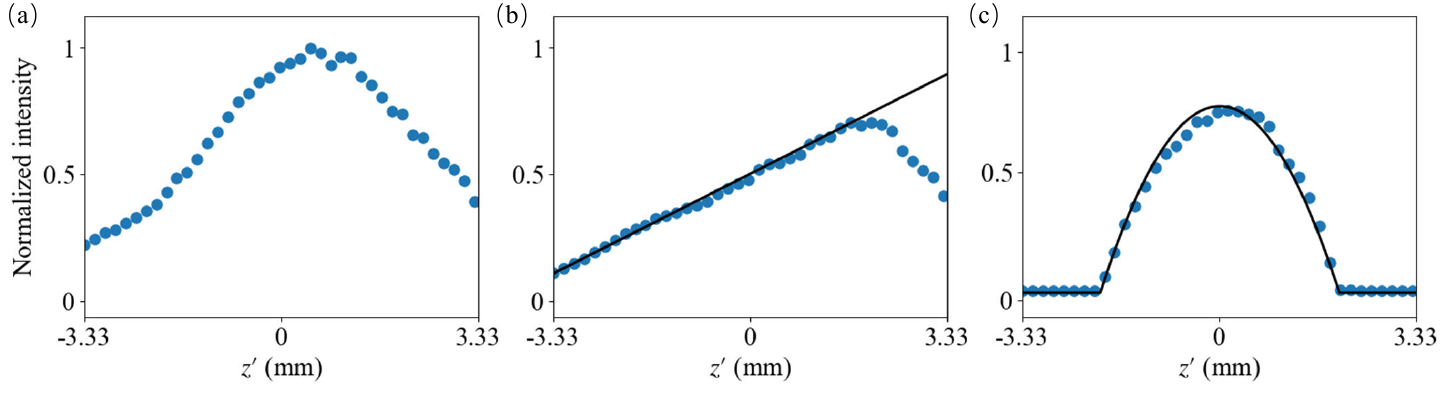}
\caption{%$\mathbf{\vert}$ \textbf{Spatial intensity modulation with spatial light modulator} 
Experimentally measured intensity distributions of (a) the Gaussian beam without modulation, (b, c) the modulated beam with linear and quadratic shapes along the $z'$ direction (blue points) and the corresponding target intensity distributions (black lines). }
\label{Fig_SM_3}
\end{figure*}

Without modulation, the spatial distribution of light incident to MOT is Gaussian-shaped as Fig.~\ref{Fig_SM_3}(a) shows. The experimentally obtained gradient and quadratic beam profiles are shown in Figs.~\ref{Fig_SM_3}(b) and (c), in agreement with the target distributions. The deviation from the linear profile in Fig.~\ref{Fig_SM_3}(b) is due to the Gaussian shape of the incident beam. In our experiments, we only shine to the MOT only the part of the beam which agrees with the target distribution.

\section{Magnetic field control with coils}

The magnetic fields in our experiment are controlled by multiple coils, including a MOT coil, three pairs of bias coils, and two pairs of quantization coils. 
The single-turn trapping coil for the 2D-MOT that generates a field with a gradient of $\SI{7}{G/ \cm}$ is switched-off $\SI{1.5}{\milli\second}$ before the storage experiments and thus has no practical effect on the storage lifetimes. 
The magnetic fields from Earth and static bias magnetic fields along all three directions are coarsely compensated by three pairs of rectangular coils of $30$, 10 and 30 turns in $x$, $y$, and $z$ directions, respectively. The longitudinal direction of MOT is defined as the $z$ axis, the horizontal direction perpendicular to the $z$ axis as $x$ axis, and the vertical direction as $y$ axis. The dimensions of these coils aligned with the $xyz$ frame are $\SI{150}{\centi\meter}\times\SI{40}{\centi\meter}\times\SI{150}{\centi\meter}$ and the generated bias fields at the center are $\left(B_x,B_y,B_z\right) =\left(\SI{64}{mG},\SI{149}{mG},\SI{64}{mG}\right)$ per $\SI{}{\A}$. 
Because of the large size of the coils compared to the geometry of the 2D-MOT ($L\approx\SI{2.5}{\centi\meter}$) ,
the gradient and curvature magnetic fields generated by the bias coils along the $z-$axis are negligible. Thus, the $B_{0,ext}$ in $z$ directions is controlled by the bias coils.

Two pairs of quantization coils with radius of $\SI{14}{\centi\meter}$ are placed with the plane of the coils perpendicular to the quantization axis and symmetrically arranged with respect to the center of the atomic ensemble. One pair consists of two single-turn coils with a distance of $\SI{14.5}{\centi\meter}$ between them, while the other pair consists of two coils with 30 turns and a distance of $\SI{12}{\centi\meter}$ between them. In the demonstration of influence of gradient and curvature of magnetic field, a Raman storage scheme is applied and the pair of multi-turn quantization coils in Helmholtz configuration generates 6.6 G magnetic field at the center of MOT for a current of 2A. The gradient magnetic field is generated by the pair of single-turn coils in anti-Helmholtz configuration with a gradient of $B_{1,ext} = 8$ mG/cm with 1.5A current as unfilled blue squares show in Fig.~3(b). With the reverse current, the storage curve is presented as unfilled triangles. In Fig.~3(c), the orange points are measured under the condition that $B_1$ is compensated by gradient magnetic field from the pair of single-turn coils as mentioned before and $B_2$ is compensated with the modulated AC Stark beam. The effective compensated curvature is $B_{2,ext} = 5.4$ \SI{}{mG/cm^2}.

The time varied magnetic field is generated by the pair of single-turn quantization coils in Helmholtz configuration which generate a magnetic field $B_{0,ext} = 60 \mathrm{mG/A}$ along the $z$ axis. The switching off time of the single-turn coils is around \SI{10}{\micro\second}.

\section{Lifetime and limitations of light pulse storage}

\begin{figure*}[htbp]
  \centering
\includegraphics[width=0.6\linewidth]{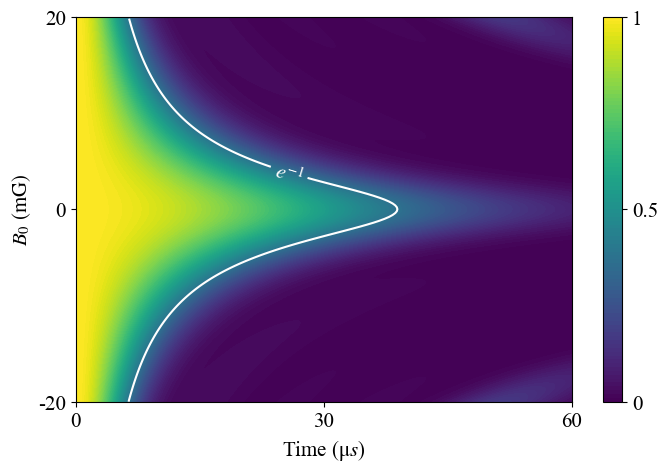}
\caption{ Normalized intensity of the retrieved light pulse as a function of storage time with different values of the magnetic field $B_0$.}
\label{FigStoragebias}
\end{figure*}

The intensity of the retrieved pulse is the superposition of the evolution of all involved $m_F$ states as Eq. 2 illustrates. To simplify and study the lifetime of storage, we initially assume that the gradient and quadratic field coefficients are $B_1 = 0$ and $B_2 = 0$. The simplified retrieved intensity is represented as follows~\cite{Jenkins2006Theory,Peters2009Optimizing}:
\begin{equation}
\eta(\tau) = |a_1 \exp{(-i 2\omega_L \tau)} + a_2\exp{(-i \omega_L \tau)} + a_3 + a_4\exp{(i \omega_L \tau)} + a_5 \exp{(i 2\omega_L \tau)}|^2\exp{(-\tau^2/T^2)},
\end{equation}
where the $a_1$ to $a_5$ present the relative population weight of each Zeeman sublevel of the $|1\rangle$ manifold, $\omega_L = \frac{2}{3}\mu_B B/\hbar$ denotes the relative phase shift. Considering the inhomogeneous atomic velocity distribution, the Gaussian decay term is added phenomenologically~\cite{Zhao2009Millisecond}. With the approximation that the temperature of atoms is low enough and the effect of decoherence is negligible, the decay term can also be presented with exponential decay~\cite{Hsiao2018Highly}, which we applied for fitting of Fig.~2(b). The normalized retrieved intensity of optical pulse under different bias magnetic fields as a function of time is shown in Fig.~\ref{FigStoragebias}. Due to the interference of the five terms, the retrieved intensity presents damped oscillation especially for strong $B_0$. The storage lifetime is defined as the time where the normalized retrieved intensity first drops to $1/e$ as the white line shows.

The storage lifetime is affected by spatial magnetic inhomogeneities such as gradient and curvature. With the Raman storage scheme, the two photon transitions of different $m_F$ levels are not degenerate. Thus, only one $m_F$ state is involved. Considering the spatial distribution, the theoretically retrieved normalized intensity is modeled by:
\begin{equation}\label{efficiencyRaman}
\eta(\tau) = |\int a_i(z) \exp{[-i \mu_B g_L m_F ( B_1 z + B_2 z^2)  ]} \text{d} z|^2 \exp{(-\tau^2/T^2)},
\end{equation}
where $a_i(z)$ is the normalized atomic density distribution of a certain $m_F$ state with $ \int a_i(z) \text{d} z = 1$. The spatially independent phase term proportional to $B_0$ does not affect the storage lifetime. The theoretical curve in Fig.~3 is predicted with Eq.~\ref{efficiencyRaman}.

\end{document}